\documentclass{article}
\usepackage[utf8]{inputenc}

\usepackage{graphicx}

\usepackage{amsmath}

\usepackage{listings}
\title{Overview of Bachelors Theses 2021}
\author{
Vitaly Aksenov, ITMO University \\
aksenov.vitaly@gmail.com
}
\date{July 2021}

\begin{document}

\maketitle

\section{Development of a Streaming Algorithm for the Decomposition of Graph Metrics to Tree Metrics}

\noindent\textbf{Student: } Fafurin Oleg, ITMO University

\noindent\textbf{External Supervisor:} Michael Kapralov, EPFL

The embedding problem.
We are given a graph $G$. We want to embed this graph onto some tree $T$, so that the shortest distance $d_G(u, v)$ between any pair of vertices $u$ and $v$ does not change much. In other words, we want to minimize $\max\limits_{(u, v)} \frac{d_T(u, v)}{d_G(u, v)}$. This value is named the \emph{distortion}. Obviously, the distortion is upper bounded by the maximal distortion of edges.

There exists an algorithm that embeds any graph on a tree with distortion $O(\log^2 n)$ in the streaming model, i.e., it can use only $O(n \cdot \mathrm{polylog}\,n)$ memory. It consists of two parts.

In the first part, we insert edges one by one and if for a given edge $(u, v)$ the current distance is less than $t$ then we do not insert it. This algorithm, obviously, provides a distortion $O(t)$ for each edge and it can be proven that the total number of edges will not exceed $O(n^{1 + \frac{1}{t}})$~\cite{greedy-spanners}. Taking $t = \log n$, we get $O(\log n)$ distortion and $O(n \cdot \mathrm{polylog}\,n)$.

In the second part, we use a streaming algorithm named \textbf{FRT}~\cite{frt}, it takes a graph with $O(n \cdot \mathrm{polylog}\,n)$ edges and gets a tree with distortion $O(\log^2 n)$.

As the first result, we improved the distortion of this algorithm by taking $t$ to be $O(\frac{\log n}{\log\log n})$ in the first part and, thus, giving $O(\frac{\log n}{\log\log n})$ distortion with $O(n \cdot \mathrm{polylog} n)$ edges in the graph. So, in total, the algorithm gives $O(\frac{\log^2 n}{\log\log n})$ distortion.

The resulting distortion is the upper bound. We decided to find graphs for which the distortion matches that upper bound. The following two graphs satisfy.

\textbf{Regular graph.} We build a regular graph with degree $O(\frac{2 \log n}{\log\log n})$: at first, put all $n$ vertices on a cycle, and then connect each vertex with $O(\frac{\log n}{\log\log n})$ neighbours in both sides.

\begin{center}
\includegraphics[scale=0.3]{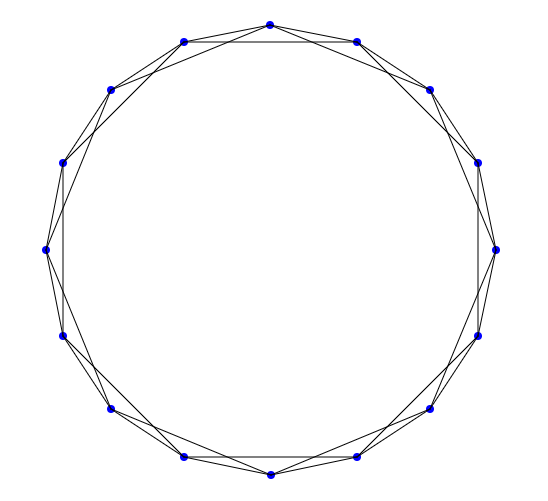}
\end{center}

\textbf{Star.} Consider $0 < \alpha < 1$. One of the vertices is a center, from which there are $n^{\alpha}$ chains with length $n^{1 - \alpha}$. Then, we take all the vertices on the distance at most $O(\frac{\log n}{\log\log n})$ from the center and add all the edges between them.

\begin{center}
\includegraphics[scale=0.3]{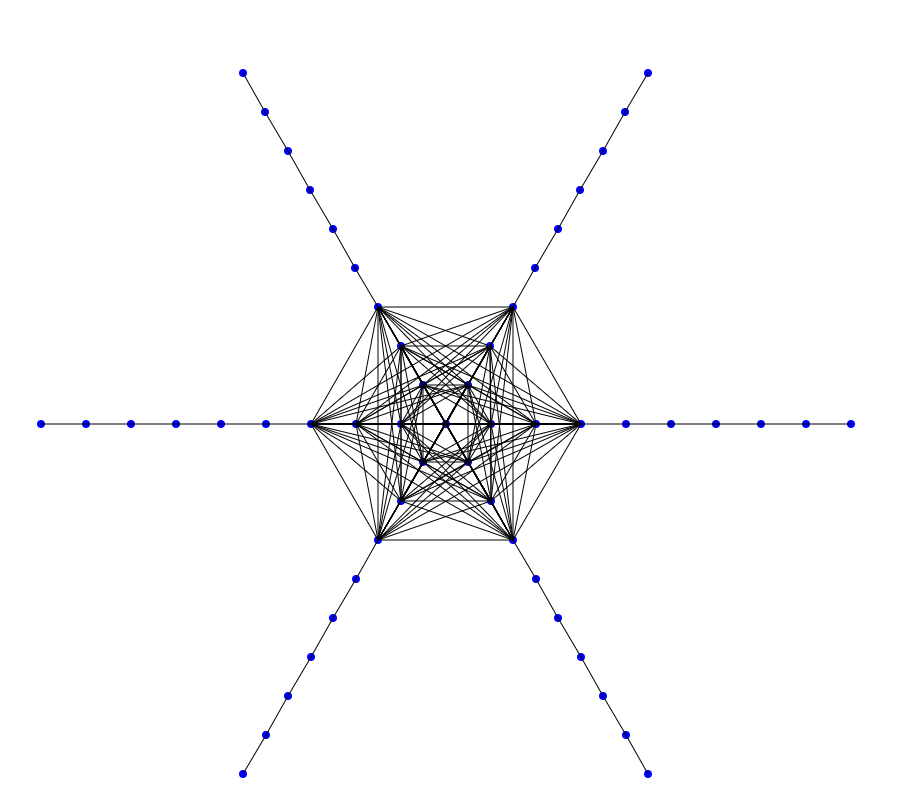}
\end{center}

Then, we implement the algorithm. The complexity of the first part appeared to be $O(mn\log n)$ where $m$ is the number of edges and $n$ is the number of vertices. The complexity of the second part is $O(n^2 \log n)$.

We run the resulting algorithm on several different open-source network graphs.

The following plot shows the distortion of paths after the first part of the algorithm on different graphs such as Facebook~\cite{snap-facebook} and scale-free graphs~\cite{power-law-base-bounds}. 

\begin{center}
\includegraphics[scale=0.4]{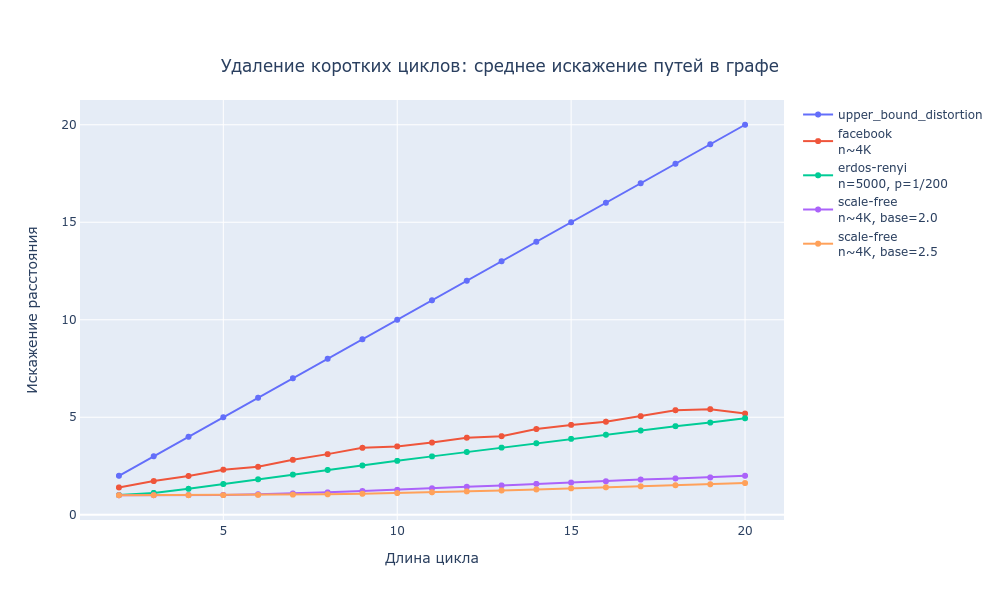}
\end{center}

The following plot shows the distortion of edges after the second part of the algorithm (FRT) on different scale-free graphs with different base.

\begin{center}
\includegraphics[scale=0.4]{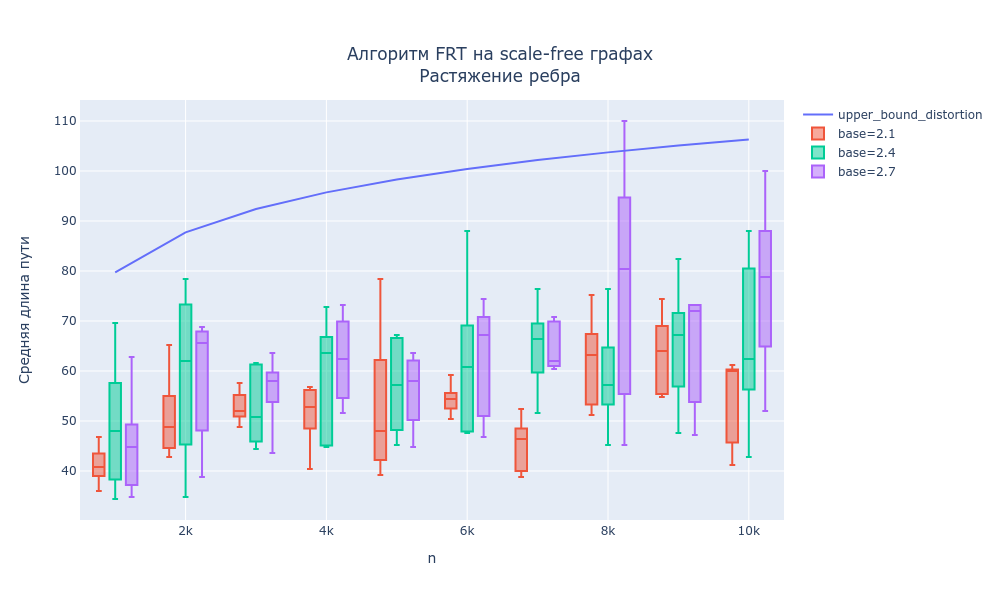}
\end{center}

\pagebreak

\section{Development of Memory-friendly Concurrent Data Structures}

\noindent\textbf{Student: } Roman Smirnov, ITMO University

\noindent\textbf{External Supervisor:} Petr Kuznetsov, Telecom Paris

The main idea of this work is to implement the skip-list so that each node can store up to $k$ elements instead of one. We designed and implemented the algorithm using locks. This thesis is mostly technical and the main results are the experiments.

At first, we chose the best $k$---it appeared to be $32$.
Then we compared our approach with two well-known concurrent data structures based on the skip-list: ConcurrentSkipListSet~\cite{java-concurrent-skip-list} from Java standard library and NonBlockingFriendlySkipListSet~\cite{contention-friendly-list}. Please, note, that we compared sets and not maps. It can be seen as that our approach does not lose the performance much.

\begin{center}
\includegraphics[width=\textwidth]{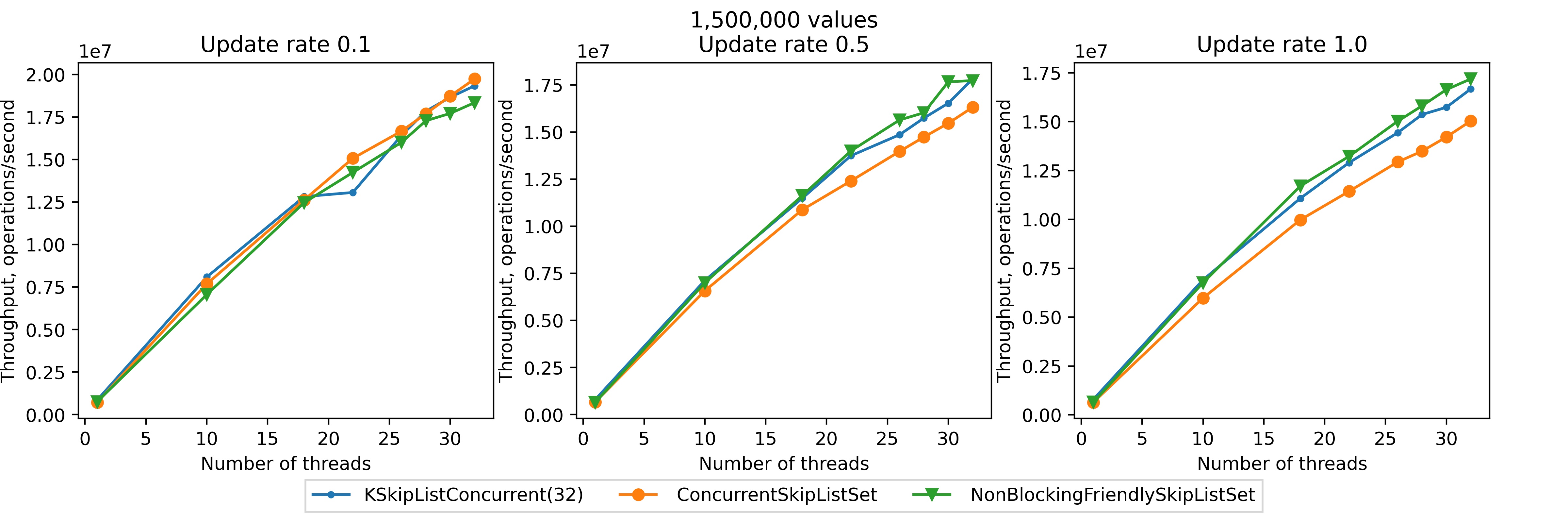}
\end{center}

Then, we decided to replace Objects in the previous implementation by integers. For that, we rewrote our algorithm and ConcurrentSkipListSet. This improved the performance of our data structure almost $2$ times since now $k$ elements reside on the same cache line, while the results of ConcurrentSkipListSet barely changed.

\begin{center}
\includegraphics[width=\textwidth]{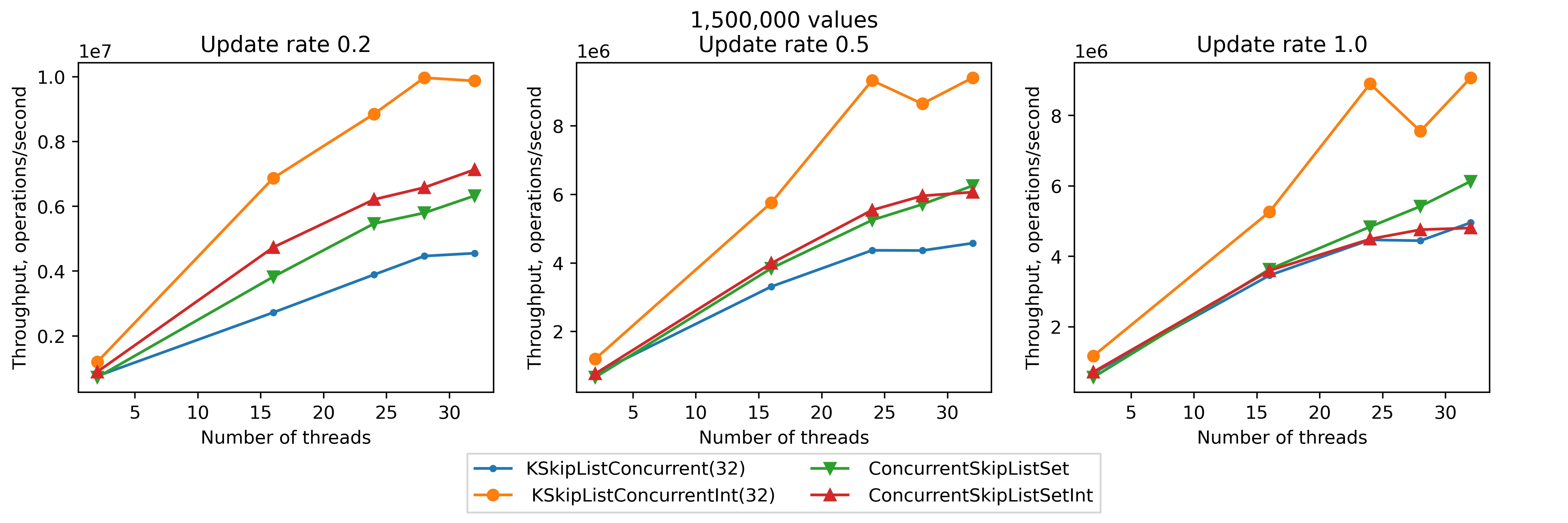}
\end{center}

As the result, we can say that the idea of batching the elements from different nodes into one seems to be a reasonable approach.

\pagebreak

\section{Theoretical Analysis of the Performance of Concurrent Data Structures}

\noindent\textbf{Student: } Daniil Bolotov, ITMO University

\noindent\textbf{External Supervisor:} Petr Kuznetsov, Telecom Paris

In this work we tried to predict the performance of MCS lock~\cite{mellor1991algorithms} and Treiber stack~\cite{treiber1986systems}. The prediction is done in the similar manner as in~\cite{aksenov2018brief}.

For MCS lock, we consider a data structure that takes MCS lock, perform the critical section of size $C$, releases the lock, and then perform the parallel section of size $P$. Thus, we can get the following code that emulates such data structure.

\begin{lstlisting}
class Node:
  bool locked // shared, atomic
  Node next = null 
  
  tail = null // shared, global
  threadlocal myNode = null // per process
operation():
  myNode = Node()                                     
  myNode.locked = true                                      |\label{line:start}|
  pred = tail.getAndSet(myNode) // $W$ or $X$  |\label{line:swap}|
  if pred != null:                                                       |\label{line:pred:condition}|
    pred.next = myNode                                        |\label{line:queue}|
    while myNode.locked: // pass // $R_I$ |\label{line:wait-critical}|
 // CS started
 for i in 1..C: // $C$                                         |\label{line:critical-start}|
    //nop                                                       |\label{line:critical-finish}|
 // CS finished  
 if myNode.next == null: // $R_I$                 |\label{line:release-1}|
   if tail.CAS(myNode,null): // $W$ or $X$         |\label{line:cas}|
     return
   else:
     while myNode.next == null: // $R_I$      |\label{line:wait-node}|
       //pass
 myNode.next.locked = false // $W$       |\label{line:unlocked-next}|
 //Parallel section
 for i in 1..P: // $P$                                       |\label{line:parallel-1}|
   //nop                                                     |\label{line:parallel-2}|
\end{lstlisting}

By considering different schedules we can prove that the throughput is equal to:
$$
\begin{cases}
\frac{\alpha}{2R_I + C + 2W} & \text{, if } P +W \leq (N - 1) \cdot (2W+C+R_I) \\
\frac{\alpha \cdot N}{(2W+C+R_I )+(P+W)} & \text{, else}
\end{cases},
$$
where $C$ is the size of the critical section, $P$ is the size of the parallel section, $W$ is the cost of a write, $R_I$ is the cost of a read, and $N$ is the number of processes.

On Intel Xeon and $15$ processes we get the following throughput, where red is the prediction and blue is the real execution:

\includegraphics[width=\linewidth]{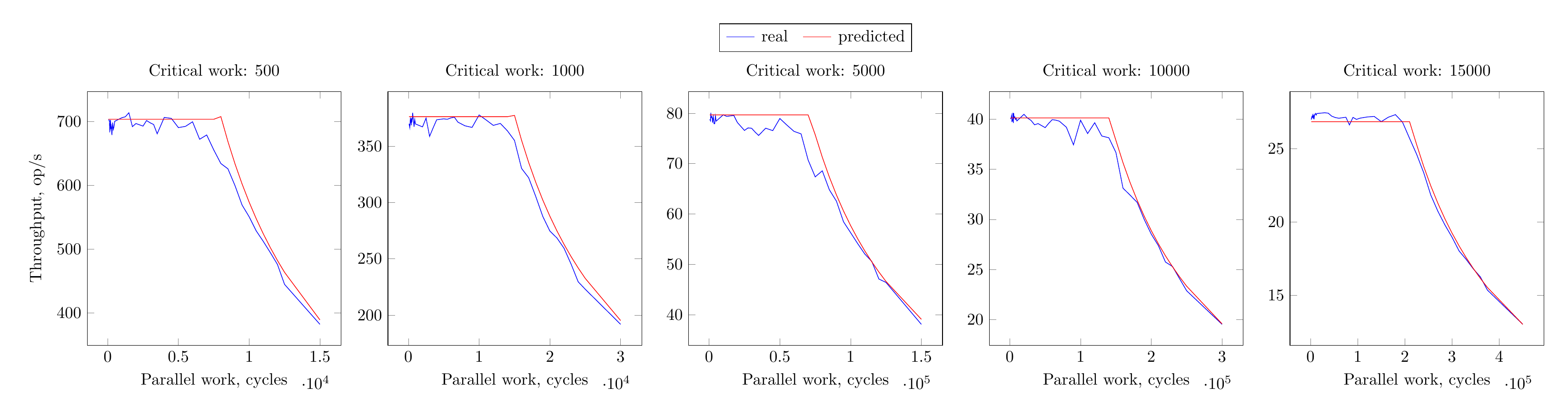}

On AMD Opteron and $15$ processes we get the following throughput:

\includegraphics[width=\linewidth]{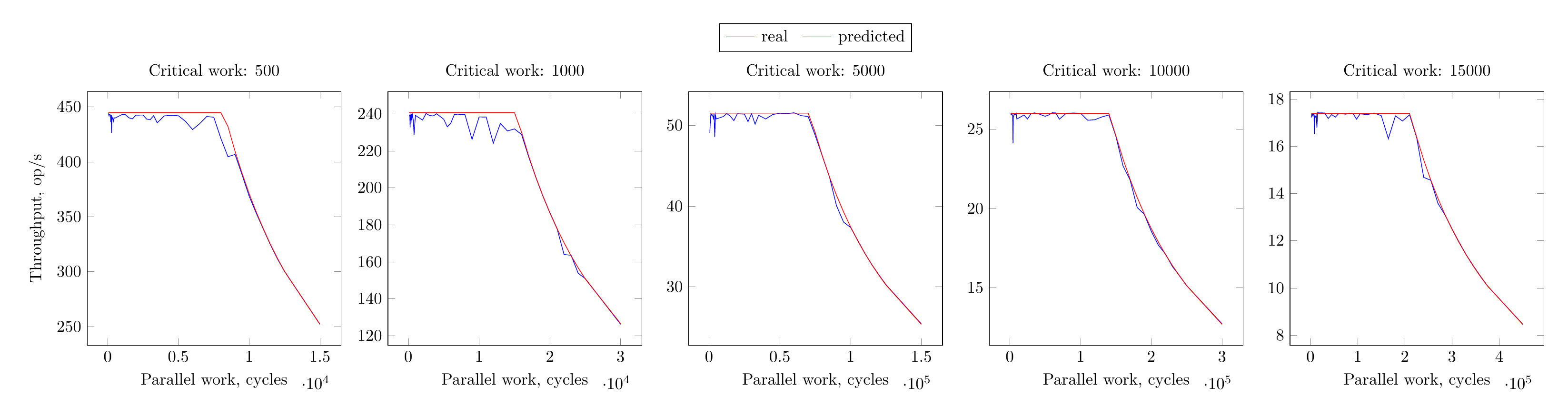}

Now, we consider Treiber stack. The pseudocode is the following:

\begin{lstlisting}
class Node:
  T data;
  Node next
  
head = null //shared, atomic
  
push(data):
  newHead = Node(data)
  while !success:
    oldHead = atomic_read(head) // $M$ or $X$
  	newHead.next = oldHead
    success = head.compareAndSet(oldHead, newHead) // $W$

pop():
  Node oldHead
  while !success:
     oldHead = atomic_read(head) // $M$ or $X$
     if (oldHead == null) {
           return DEFAULT_VALUE // $\text{corner case}$
     }
     newHead = oldHead.next
     success = head.compareAndSet(oldHead, newHead) // $W$
     
  return oldHead.data
\end{lstlisting}

One can see that \texttt{push} and \texttt{pop} operations are similar and we can write them as one generic function as follows:
\begin{lstlisting}
pop_or_push_operation():
  while !success do                             
    current = atomic_read(head)                 |\label{line:retry:start}|
    new = critical_work(current)
    success = head.compareAndSet(current, new)    |\label{line:retry:end}|
\end{lstlisting}

Then, we simulate the application of the Treiber stack: we take an element from the stack and then we perform an execution of size $P$.

\begin{lstlisting}
class Node:
  T data;
  Node next
  
head = null //shared, atomic
  
operation():
  newHead = Node(data)
  while !success:
    oldHead = atomic_read(head) // $M$ or $X$  |\label{line:atomic_load}|
  	newHead.next = oldHead                   |\label{line:um}|
    success = head.compareAndSet(oldHead, newHead); // $W$ |\label{line:cass}|
    
  for i in 1..P: / / $P$                                       |\label{line:trparallel-1}|
    //nop                                                     |\label{line:trparallel-2}|
\end{lstlisting}

By considering different schedules we can prove that the throughput is equal to:
$$
\begin{cases}
\frac{\alpha}{M+W} & \text{, if } P \leq (N - 1) \cdot (M + W) \\
\frac{\alpha \cdot N}{(P+M+W)} & \text{, else}
\end{cases}
$$

On Intel Xeon and $15$ processes we get the following results:

\includegraphics[width=\linewidth]{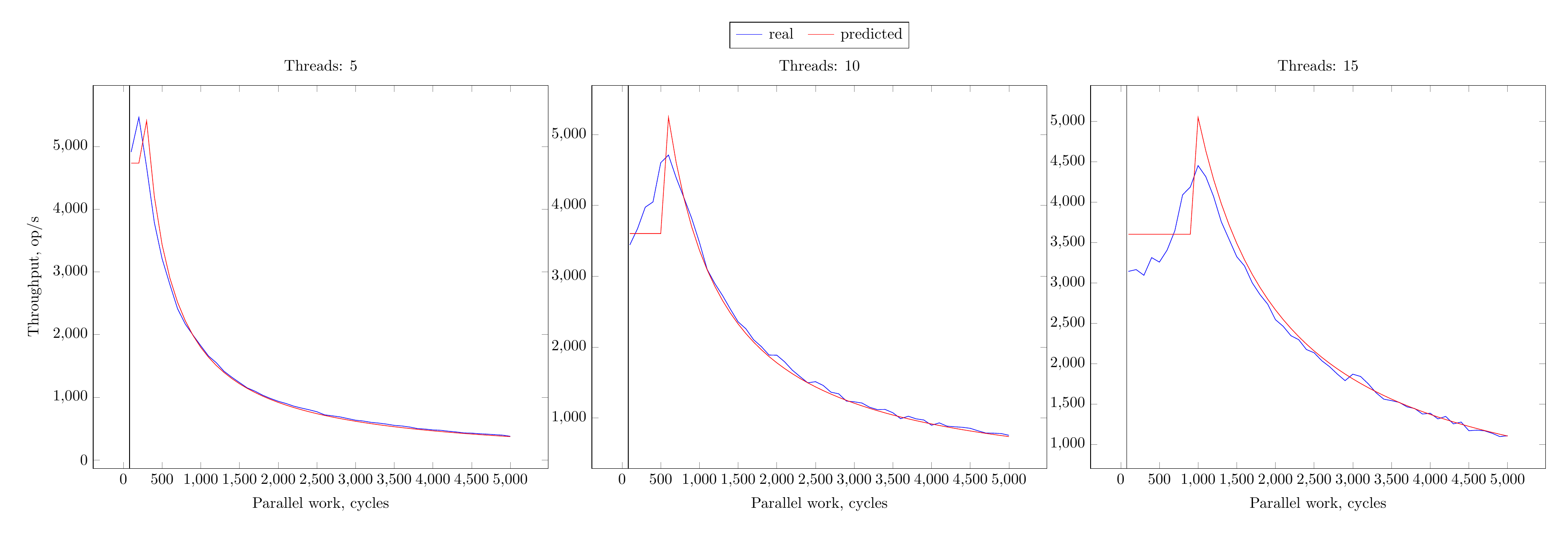}

On AMD Opteron and $15$ processes we get the following results:

\includegraphics[width=\linewidth]{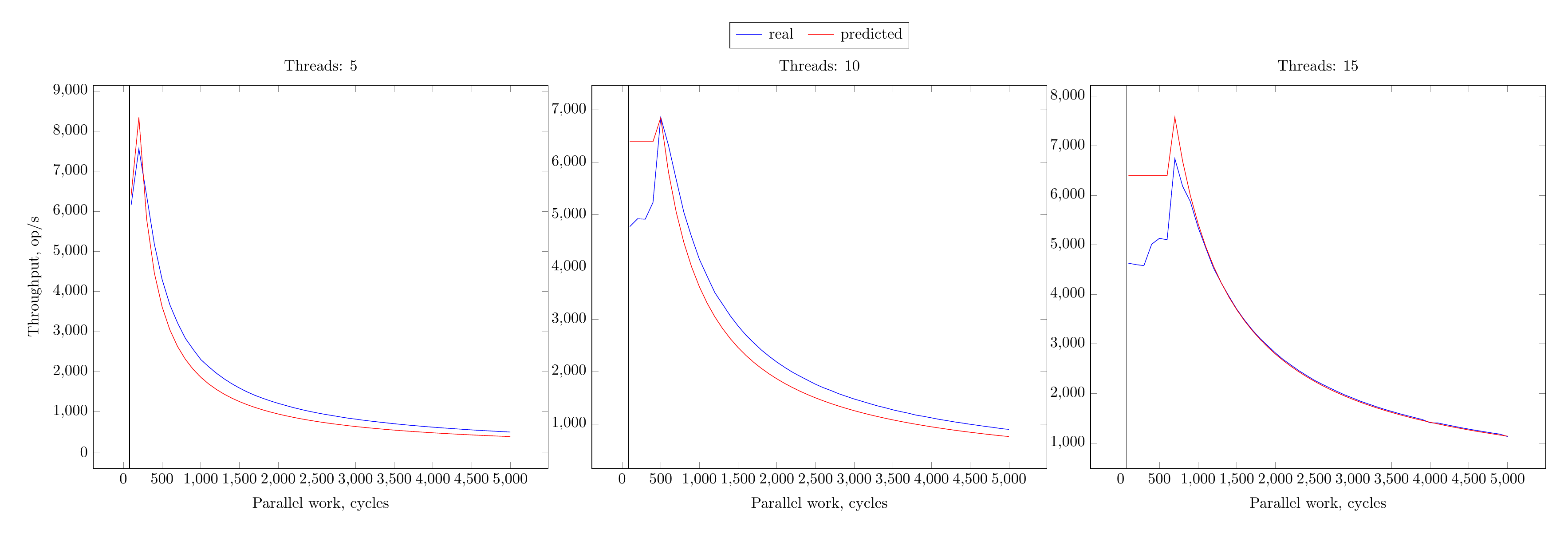}

As a result, we get pretty good theoretical approximation of the throughput.

\pagebreak

\section{Parallel Batched Interpolation Search Tree}

\noindent\textbf{Student:} Alena Martsenyuk, MIPT

In this thesis, we show how to design \emph{parallel batched} implementation of Interpolation Search Tree~\cite{mehlhorn1985dynamic}. ``Parallel batched'' means that we ask the data structure to apply multiple operations together in parallel.

We developed the data structure that applies a batch of $m$ operations in $O(m \log\log n)$ work and $O(\log m \log\log n)$ span, where $n$ is the current size of the tree.

For experiments, we used an Intel Xeon machine with $16$ threads.
On this plot, you can see how much time (OY-axis) it takes to apply $m$ (OX-axis) operations using different number of processes into a tree of size $2.5 \cdot 10^7$.

\includegraphics[width=\textwidth]{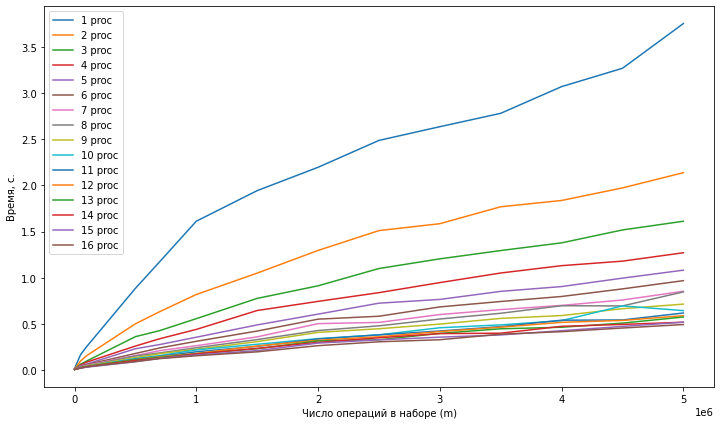}

On this plot, you can see how much time (OY-axis) it takes to apply $10^6$ operations using different number of processes into a tree of size $n$ (OX-axis).

\includegraphics[width=\textwidth]{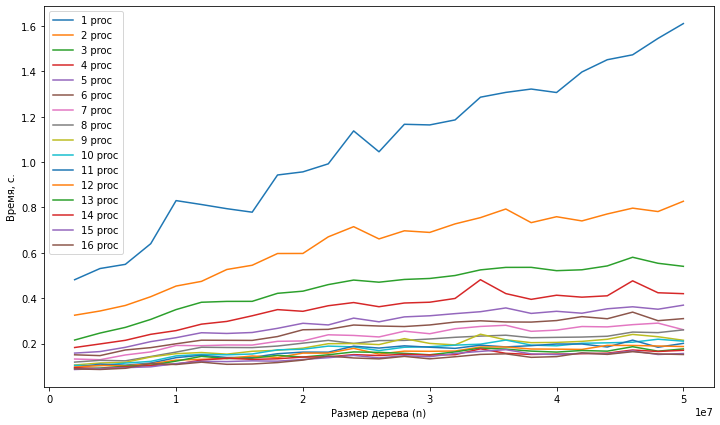}

Finally, we insert $10^6$ elements into the tree of size $5 \cdot 10^7$ and check the speedup. The speedup is approximately $11$ on $16$ processes. 

\includegraphics[width=\textwidth]{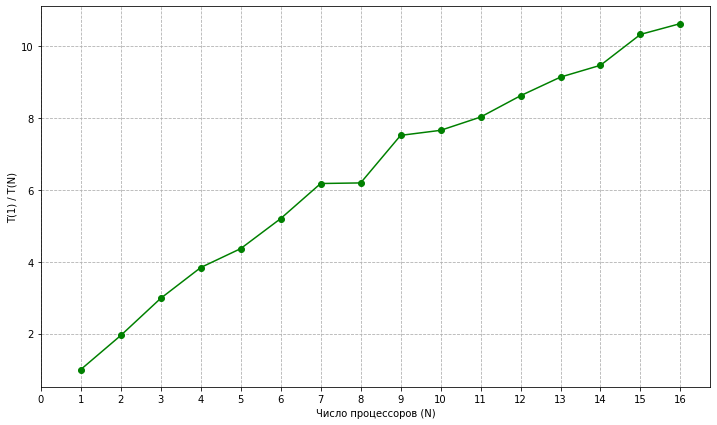}

\pagebreak

\section{Parallel Batched Self-adjusting Data Structures}

\noindent\textbf{Student:} Vitalii Krasnov, MIPT

In this thesis, we show how to design parallel batched self-adjusting binary search tree. We based our data structure on CBTree data structure~\cite{Afek2012}.

We proved that the resulting data structure is static-optimal, i.e., the total work is equal to $O(\sum\limits_x c_x \cdot \frac{m}{c_x})$ where $m$ is the total number of operations from the start of the existence of the data structure and $c_x$ is the number of times $x$ is requested. The span of the algorithm is $\frac{m}{C}$ where $C$ is $\min\limits_x c_x$.

For experiments, we used an Intel Xeon machine with $16$ threads. All our experiments has the following construction: we continuously add $10^3$ elements to the same tree until it becomes very large---so, the tree is always the same but growing.
On the first plot, one can see how much time (OY-axis) it takes to apply batches of size $10^3$ into a growing tree (OX-axis). The speedup is approximately $9$ on $12$ processes.

\includegraphics[width=\textwidth]{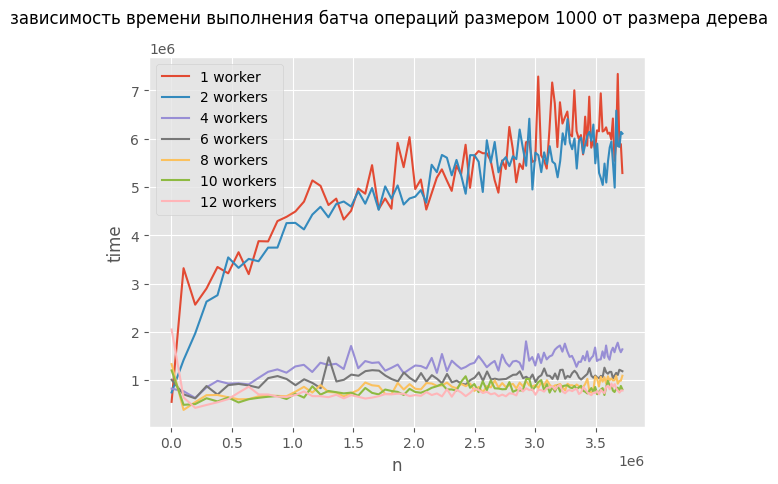}

On the second plot, one can see how much time (OY-axis) it takes to apply batches of size $10^3$ taken from a normal distribution into a growing tree (OX-axis).

\includegraphics[width=\textwidth]{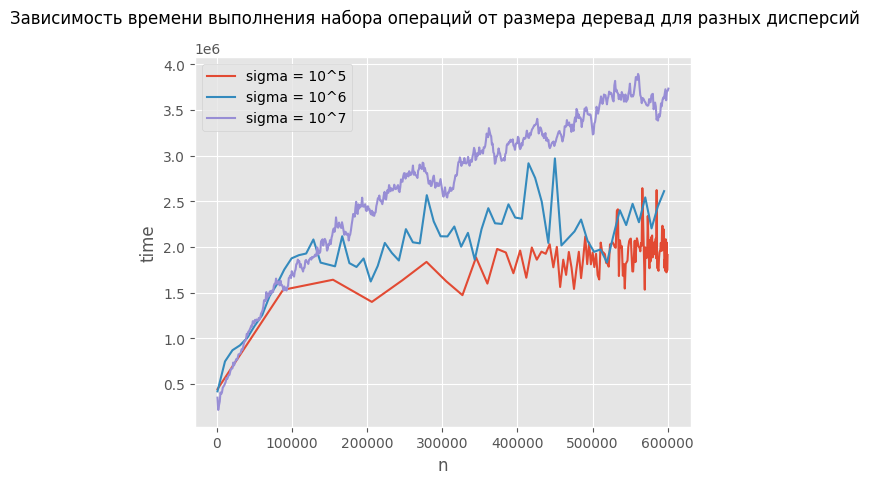}
 
Also, our data structure outperforms the set data structure from the standard C++ library in the sequential setting.
 
\includegraphics[width=\textwidth]{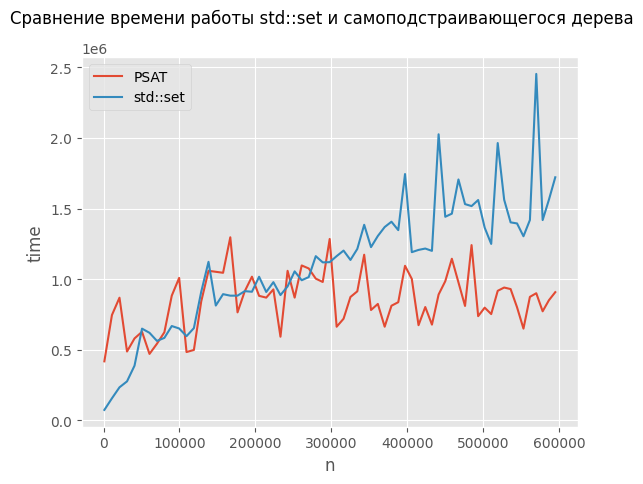}
 
 \pagebreak
 
\section{Parallel Batched Persistent Binary Search Trees}

\noindent\textbf{Student:} Ildar Zinatulin, MIPT

In this thesis, we show how to design a persistent parallel batched binary search tree. We consider persistence in the sense of versions. Suppose we are asked to apply operations $op_1, op_2, \ldots, op_m$. A result of any operation is the new version of the tree, and operations should be applied in some ``sequential'' order $op_{\pi(1)}, \ldots, op_{\pi(m)}$, i.e., a version of the tree after operation $op_{\pi(j)}$ should be the initial tree after an application of all first $j$ operations  $op_{\pi(1)}, \ldots, op_{\pi(j)}$.

We designed a persistent binary search tree that applies the operations in the order of their arguments. The idea is a little bit complicated and is similar to the scan function~--- we make two traversals from top to bottom. The work of the resulting algorithm is $O(m \log n)$ and the span is $O(\log n \log m)$.

For experiments, we used an Intel Xeon machine with $16$ threads. We performed only one experiment~--- the speedup of an application of a batch with size $10^5$ to a tree with size $10^6$. As for the binary search tree we used Treap. The blue dot on the plot is the sequential algorithm for the persistent Treap.

\includegraphics[width=\textwidth]{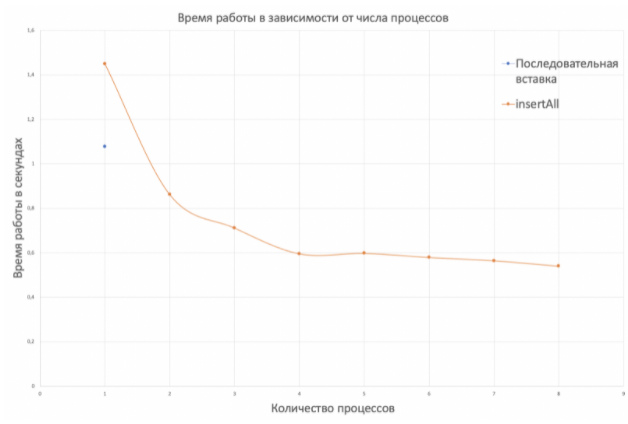}

\pagebreak

\bibliographystyle{abbrv}

\bibliography{references}

\end{document}